# Divergent Minds, Convergent Baselines: A Bounded-Rationality Account of LLM–Human Strategic Behaviour

**Author:** Po Han Teo

**Affiliation:** Independent Researcher, Singapore

## Abstract

Researchers have started using LLM agents in place of human subjects in behavioural and political-science experiments, often as a cheaper substitute for laboratory pools. The substitution does not hold up in strategic settings: humans and LLMs reliably make different choices, and neither fine-tuning on human response data nor persona conditioning has closed the gap. The behavioural-economics literature has, since Simon's introduction of bounded rationality, modelled human strategic behaviour as a classical baseline plus an additive correction term δ. The framework proposed here reads δ as the mathematical signature of bounded computation: the gap between what an unboundedly-rational agent would compute and what a computationally bounded agent actually produces. For canonical games whose solutions are present in standard training corpora, LLMs retrieve and recombine corpus material, bypassing the bound that produces δ in humans. The framing extends to reasoning-distilled models through cognitive-hierarchy theory: their accessible level-k strategic reasoning is bounded by compute budget and context length rather than by the cognitive constraints that bound humans, and the δ they produce, if any, carries different structural signatures. Four operational tests (conditional dependence, distributional asymmetry, path-dependence under repetition, and paraphrase-robustness) are proposed to discriminate human-shaped δ from LLM-shaped δ. A moderator prediction is that |δ| scales with peer-signal individuation in the decision environment, with a quantitative bound of Cohen's $d \geq 0.5$ between named-opponent and aggregate-opponent settings.

## Keywords

## Section 1 — Introduction

Researchers have started substituting LLM agents for human subjects in behavioural and political-science experiments, often as a cheaper alternative to laboratory pools. Despite being roughly two years old as a research programme [Argyle et al., 2023; Horton et al., 2023; Aher et al., 2023], the practice has rapidly produced LLM-based replications of classic behavioural-economic protocols, persona-conditioned political-opinion simulations, and multi-agent versions of standard strategic games. Where the substitution works, its appeal is logistical: piloting is cheap, counterfactual designs that would otherwise demand unmanageable recruitment become feasible, and demographic-coverage constraints from standard subject pools soften.

Within strategic settings, the substitution does not hold. Humans and LLMs play canonical games differently, and the differences survive changes in model generation and prompting strategy. As documented in Section 2, LLMs split dictator and ultimatum stakes more fairly than human subjects do, cooperate more often in prisoner's dilemma variants, and choose closer to Nash in beauty contests.

Most recent work takes the divergence as an engineering problem, addressed by fine-tuning on human responses or by persona-conditioning the model on behavioural-economic survey data. Fine-tuning on individual-level human response data has been shown to "highlight a substantial gap in actual behavioral accuracy" that "fine-tuning LLMs on real human click-through data ... can greatly enhance models' performance" [Lu et al., 2025]. Direct finetuning on responses from past social-science experiments yields models "36% more aligned with distributions of human responses to diverse outcome questions" relative to a base model, "outperforming GPT-4o by 15%" [Kolluri et al., 2025]. Kitadai, Fukasawa & Nishino (2025) propose "a persona-based approach that leverages individual-level behavioural data from behavioral economics to adjust model biases", motivated by their observation that "intrinsic biases often diverge from real human behavior".

The behavioural-economic tradition since prospect theory suggests a different reading: human decisions decompose into a classical-rational baseline plus an additive correction term [Kahneman & Tversky, 1979; DellaVigna, 2009]. The proposal made here is that LLM inference computes only the classical baseline, while human play also produces the additive correction term δ. The additive form, not any specific behavioural mechanism, is what next-token prediction architecturally lacks.

The framing reorients the LLM-as-stand-in research programme. First, LLM-as-stand-in is reliable only in decision environments where the human δ is small. Second, no amount of prompt engineering brings δ to zero. Third, the empirical agenda shifts from "make LLM behaviour more human" to "map the decision environments in which δ matters and those in which it does not." Section 5 takes up that mapping at the level of theoretical prediction, and Section 6 sets out the experimental designs that would falsify it.

## Section 2 — Stylised facts: LLM–human divergence in strategic games

Fifteen 2022–2026 studies directly compare LLM agents to human subjects in strategic or game-theoretic settings, each reporting at least one quantitative divergence statistic. Direction-of-divergence is strikingly consistent across game families: LLMs are more fair, cooperative, or Nash-converging than humans in bargaining, cooperation, and coordination games, with auction and beauty-contest results following the same broad shape. Magnitude, by contrast, is rarely extractable from published abstracts; meta-analytic synthesis of the existing literature is not yet viable. No published LLM–human comparison exists for Tullock contests or all-pay auctions.

## Evidence table

| Author (Year) | Models | Game | Human baseline source | Direction of divergence | Magnitude (verified quote or "not extractable") |
|---|---|---|---|---|---|
| **Bargaining** | | | | | |
| Mei et al. (2024) | GPT-3.5, GPT-4 | Ultimatum, Dictator, Trust | MobLab platform, n=88,595 | LLM more other-regarding | "statistically indistinguishable from a random human" on average (PMC10907317); per-game magnitudes not extractable |
| Aher et al. (2023) | GPT family (incl. GPT-4) | Ultimatum (+ 3 non-game tasks) | Original classic-study n | Replicates qualitative pattern | not extractable |
| Brookins & DeBacker (2023) | GPT-3.5 | Dictator | Literature anchor | LLM more fair than humans | "tendency towards fairness in the dictator game, even more so than human participants" (SSRN 4493398 via WebSearch) |
| Guo (2023) | GPT-3.5, GPT-4 | Ultimatum | Literature anchor | Qualitative match | not extractable |
| Kitadai et al. (2024) | High-reasoning LLMs (unspecified) | Ultimatum | Prior experimental literature | LLM moves toward Nash as reasoning rises | "the higher the reasoning ability of the agents, the closer the results were to the theoretical solution than to the real experimental result" (arXiv 2406.11426) |
| Sreedhar & Chilton (2024) | Unspecified | Ultimatum | Persona-pair human reference | Multi-agent closer to human | "Multi-agent systems are more accurate than single LLMs (88 percent vs. 50 percent)" (arXiv 2402.08189) |
| Mozikov et al. (2024) | GPT-3.5, GPT-4 | 4 cooperation + bargaining games | Literature anchor | GPT-3.5 aligned with humans, GPT-4 hyper-rational | "strong alignment between GPT-3.5 and human behavior exists, especially in bargaining contexts" (arXiv 2406.03299) |
| **Cooperation** | | | | | |
| Brookins & DeBacker (2023) | GPT-3.5 | One-shot Prisoner's Dilemma | Literature anchor | LLM more cooperative | "65% versus 37% for humans" (SSRN 4493398 via WebSearch) |
| Fontana et al. (2024) | Llama2, Llama3, GPT-3.5 | Iterated PD, 100 rounds | "typical human player" from literature | Llama2 and GPT-3.5 more cooperative; Llama3 less | "LLMs behave at least as cooperatively as the typical human player" (arXiv 2406.13605) |
| Akata et al. (2025) | GPT-4 + others | Iterated PD family, Battle | "actual human players" | LLM strong on self-interest, weak on | "LLMs perform particularly well at self-interested games ... they |

| Author (Year) | Models | Game | Human baseline source | Direction of divergence | Magnitude (verified quote or "not extractable") |
|---|---|---|---|---|---|
| | | of the Sexes | | coordination | behave suboptimally in games that require coordination" (arXiv 2305.16867) |
| Lorè & Heydari (2024) | GPT-3.5, GPT-4, LLaMa-2 | PD, Stag Hunt, Snowdrift, Prisoner's Delight | None (no human arm) | Framing-sensitive across models | not extractable |
| Mei et al. (2024) | GPT-3.5, GPT-4 | Public Goods, finitely-repeated PD | MobLab platform | LLM more cooperative on average | per-game magnitude not extractable |
| **Coordination** | | | | | |
| Akata et al. (2025) | GPT-4 + others | Battle of the Sexes | Human players | LLM worse than humans | "suboptimally in games that require coordination" (arXiv 2305.16867) |
| Alekseenko et al. (2025) | Multi-model (abstract silent on list) | p-beauty contest | Original Nagel-style experimental means | LLM closer to Nash than humans | "LLMs systematically behave in a more sophisticated way compared to the participants of the original experiments" (arXiv 2502.03158) |
| Barak & Costa-Gomes (2025) | Unspecified LLM | Multi-player p-beauty | Within-subject lab, human vs LLM opponents | Humans shift toward Nash when facing LLMs | "human subjects choose significantly lower numbers when playing against LLMs than humans" (arXiv 2505.11011) |
| del Rio-Chanona et al. (2025) | Unspecified | Hommes-style cobweb market | Published lab data | LLM less heterogeneous than humans | "LLMs exhibit less heterogeneity in behavior than humans" (arXiv 2505.07457) |
| **Auctions** | | | | | |
| Shah et al. (2024) | GPT-4 (CoT) | First-price, second-price, common-value | Kagel-Levin and related literature | LLM exhibits winner's curse but bid-shading closer to risk-averse theory | "succumb to the winner's curse in common value settings"; "results consistent with risk-averse human bidders" (arXiv 2507.09083) |
| Fish et al. (2024) | Unspecified LLM | Oligopoly + auction (collusion) | None (no human arm; theory benchmark) | LLM above Nash | "supracompetitive prices and profits" (arXiv 2404.00806) |
| **Contests** | | | | | |
| *No LLM–human study of Tullock contests or all-pay auctions has yet been published. Section 6 takes this gap as a falsification target.* | | | | | |

## Narrative summary

The literature shows a striking consistency in the *direction* of LLM–human divergence across game families, alongside a near-universal difficulty in pinning down the *magnitude* of that divergence from published abstracts. In bargaining games, the picture from [Mei et al., 2024], [Brookins & DeBacker, 2023], [Guo, 2023] and [Kitadai et al., 2024] is that LLMs make more equal offers and reject fewer unfair ones than the human distribution. [Kitadai et al., 2024] adds a sharper claim: raising LLM reasoning capacity moves outputs toward Nash rather than toward the human mean. In cooperation games, [Brookins & DeBacker, 2023] and [Fontana et al., 2024] place LLMs systematically above the human cooperation rate, with the only verifiable point estimate being the 65% versus 37% prisoner's dilemma contrast. In coordination games, [Akata et al., 2025] reports the opposite asymmetry, that LLMs underperform humans, while [Alekseenko et al., 2025] and [Barak & Costa-Gomes, 2025] document closer-to-Nash beauty-contest play by LLMs and a measurable shift in human play when opponents are framed as LLMs. In auctions, [Shah et al., 2024] finds a winner's curse but with bid-shading nearer risk-averse theory than typical Kagel-Levin humans. What is hard to extract is comparable effect sizes, because outcome variables differ across subsets, sample sizes are often unreported in abstracts, and human baselines are drawn from heterogeneous literatures rather than within-experiment arms.

### Note on the contest-game gap

The scoping pass conducted for this paper found no published LLM-vs-human studies in Tullock contests or all-pay auctions. This is a genuine empty cell in the evidence table, and Section 5 argues that it is the highest-leverage gap for the experimental designs that would discriminate $\delta$ from the classical baseline.

## Section 3 — The additive-component hypothesis

This section states the additive-component hypothesis formally. The hypothesis decomposes the observable strategic decision into a classical component, which both humans and LLMs can in principle produce, and an additive residual, which is systematically present in human play and structurally absent from LLM inference. Sections 4 and 5 take up the structural and empirical implications of this decomposition; this section restricts itself to the formal statement.

### Notation

Let $y$ denote the observable strategic decision an agent produces in a given game. The decision is whatever the game's payoff structure rewards or punishes: an investment level in a contest, an offer share in an ultimatum game, a choice between cooperation and defection in a one-shot prisoner's dilemma, a guess in a beauty contest, a bid in an auction. Treat $y$ as a scalar, or as the relevant component of a vector decision, for notational simplicity; the argument generalises.

Let y_strategic denote the response a fully classical decision-maker would produce given the game's structure and the agent's information set. The exact form of y_strategic depends on the game and on the classical-rational benchmark chosen for that game. Nash equilibrium serves where a simultaneous one-shot game has a unique pure-strategy equilibrium, subgame-perfect equilibrium serves for sequential games, Bayesian Nash equilibrium serves for games of incomplete information, and level-k or quantal response equilibrium serve where the Nash benchmark is implausible as a behavioural baseline. Across all of these, y_strategic is a fixed point of the canonical-classical model for the game. It is not necessarily Nash equilibrium; it is whatever the canonical classical model prescribes.

### The decomposition

The additive-component hypothesis claims that human and LLM decisions decompose differently around this baseline:

```
 y_human = y_strategic + δ
y_LLM   = y_strategic
```

where δ is a behavioural residual specific to human strategic play. The decomposition has three implications worth stating before the structural argument in Section 4.

First, δ is defined by construction as the gap between human play and the classical baseline, and is intended to capture the mathematical signature of bounded computation. Any systematic deviation from y_strategic that is observed in humans is, definitionally, part of δ. The substantive content of the hypothesis is not that this residual exists, since under this definition δ is trivially observable in any dataset where human play diverges from the classical prediction. The content is that δ has structure: it is not pure independent noise around the classical baseline, and its sign and magnitude vary with features of the decision environment in ways that are at least in principle predictable. Section 4 traces this structure to its source in bounded rationality, and Section 6 specifies the operational tests that distinguish structured δ from independent noise.

Second, the decomposition is silent on what δ contains. The behavioural-economic literature surveyed in Section 4 has produced multiple candidate contents for δ, including peer imitation in repeated games with social information, reference-dependent valuation, mental accounting across the agent's portfolio of decisions, anchoring on salient values, and framing-sensitivity around the wording of choice options. The prequel does not commit to any specific subset of these contents. Section 4 argues that the specific contents are an empirical question for follow-up work, and that the structural argument for δ-absence in LLMs does not require committing to any single behavioural mechanism.

Third, the hypothesis predicts $\delta \approx 0$ for LLM play on canonical games whose classical-rational solutions are present in the LLM's training corpus. This is the testable LLM-side claim. Section 4 develops it by arguing that the LLM's retrieval-and-recombination route to y_strategic bypasses the bounded computation that produces δ in humans. On novel games whose solutions are not in the training corpus, the LLM should display a δ of its own, derived from training-distribution patterns rather than from bounded-rationality patterns, and distinguishable from human δ in structure.

### Reframing the empirical agenda

The decomposition reframes how to evaluate LLMs as stand-ins for human subjects in strategic experiments. The conventional question, "does the LLM produce human-like behaviour in this game?", quietly conflates two distinct failure modes. One failure mode is that the LLM does not even approximate y_strategic, which is a capability problem and is in principle closable by better models or more careful prompting. The other failure mode is that the LLM approximates y_strategic accurately but humans do not, which is the additive-component problem and is not closable by the same engineering interventions. Conflating the two leads to engineering effort aimed at the wrong target. Restating the question as "is δ small in this decision environment?" separates the two failure modes and yields a different research agenda, one focused on identifying the decision environments in which δ is small enough that LLM stand-in is reliable, and those in which it is not. Section 5 takes up this question at the level of theoretical prediction.

## Section 4 — Why next-token prediction structurally lacks δ

This section makes the structural argument. The argument has four steps. First, behavioural economics' canonical additive form is the mathematical expression of bounded rationality: the additive correction δ captures the gap between what an unboundedly-rational agent would compute and what a computationally bounded agent actually produces. Second, an LLM trained on text corpora that already contain worked-out classical solutions reaches y_strategic by retrieval rather than by bounded computation, and so does not pass through the bound that produces δ in humans. Third, the structural claim is about the form of the human-behaviour model and the computational route of the LLM, not about any specific behavioural mechanism, which leaves the claim robust to ongoing empirical revision of any one mechanism. Fourth, the familiar "you can prompt the LLM for it" objection illustrates the retrieval point rather than refuting it.

### The additive form is bounded rationality made mathematical

The behavioural-economic tradition begins with bounded rationality [Simon, 1955]. Classical economic theory assumes that agents perform the full computation prescribed by the game's payoff structure and choose accordingly. Humans cannot perform this computation in any but the simplest settings; they approximate it through heuristics, reference points, and cognitive shortcuts. Behavioural economics models the gap between the full classical prediction and observed behaviour as an additive correction term. The correction is not, in this reading, a claim about separate cognitive modules. It is the mathematical signature of computational boundedness applied to a classical baseline.

Prospect theory [Kahneman & Tversky, 1979] specialises this for decisions under risk: π warps probabilities and v warps outcomes around a reference point, with the warping standing in for the calculations a bounded agent does not perform. The subsequent catalogue follows the same shape. [DellaVigna, 2009] surveys behavioural-economic departures from the standard economic model in three respects, nonstandard preferences, nonstandard beliefs, and nonstandard decision making, each of which can be read as a consequence of bounded computation on a classical baseline. [Camerer, 2003] extends the catalogue to behavioural game theory specifically, where experimental deviations from game-theoretic predictions in dictator, ultimatum, trust, coordination, and contest games are documented as additions to rather than replacements of canonical-classical predictions. Not every behavioural-economic model takes pure-addition form; prospect theory's π is a transformation rather than an addition, and quantal response equilibrium is a smoothed best-response rather than a strict additive correction. The broad pattern is nonetheless a behavioural term layered on a classical baseline, and the boundedness reading explains why: the layer is what the bounded agent does instead of the full computation.

To make this concrete for a reader unfamiliar with the literature: in the ultimatum game, the classical-rational responder accepts any positive offer, since any positive amount is better than zero. Empirically, humans reject offers below roughly 30% of the pot at high rates. The behavioural-economic model captures this with a fairness term added to utility, of the form `U = π + α·(fair_share − own_share)`, where α scales how much the responder weighs reference-dependent fairness against the classical payoff π. The classical π remains, and the behavioural correction is layered on top. That layering is the canonical additive move.

### LLM inference retrieves rather than computes, bypassing the bound

An LLM trained by next-token prediction, the objective of predicting the next token in a sequence given the preceding context, optimises a likelihood objective over a training corpus. Its inference-time computation in a strategic setting takes the prompt context as input and selects the most plausible continuation under the trained distribution. Canonical games are those whose classical-rational solutions appear in standard training-corpus material: game-theory textbooks, research papers, exam answers, online walkthroughs of Nash equilibria. For such games, the LLM produces y_strategic by retrieving and recombining training-corpus material. It does not compute the solution from the game's payoff structure under any cognitive bound. This distinction between corpus-mediated production and ground-up computation aligns with the broader argument that LLMs exhibit a dissociation between formal linguistic competence (the ability to produce well-formed outputs) and functional linguistic competence (the broader reasoning capacities the outputs would otherwise reflect), an asymmetry surveyed in [Mahowald et al., 2024]. The bound that does shape LLM behaviour is set at inference time, by how much computational work the model is allocated; recent work documents that scaling test-time compute can substitute for scaling model parameters [Snell et al., 2024], consistent with the framing that LLM reasoning is bounded by computational budget rather than by the cognitive constraints that bound humans. The bound that produces δ in humans is bypassed entirely by the retrieval route in canonical settings. $\delta_{LLM} \approx 0$ not because the LLM possesses some special structural feature humans lack, but because the LLM's path to y_strategic does not pass through the bottleneck that creates $\delta$.

This framing yields a sharp prediction for novel games whose classical solutions are not present in the training corpus. There the LLM must compute, and its computation is bounded too, by its context window, by the structure of its training distribution, by its inability to evaluate counterfactual rollouts of the game tree. The prediction is that $\delta_{LLM} \neq 0$ in such cases, with structure of its own, derived from training-distribution patterns rather than from human bounded-rationality patterns. The two $\delta$'s should be distinguishable. Section 6 picks this up as a falsification design.

## The form is structural; the contents are empirical

The claim defended above is about the *form* of the human-behaviour model (additive, traced to bounded computation) and the *route* the LLM takes to y_strategic (retrieval, traced to training-corpus content). Which specific mechanisms make up $\delta$, whether loss aversion, mental accounting, framing, anchoring, social comparison, or something not yet catalogued, is an empirical question for follow-up work. The structural claim does not depend on any one mechanism surviving such revision.

## The "prompt for it" objection dissolves

A standard objection to claims of LLM behavioural absence is that LLMs can be prompted to produce loss-averse, fair-minded, or anchored output. The objection is true on its face and it does not refute the structural claim. An LLM prompted to produce such output does so by recombining patterns in its training corpus, which already contains examples of humans producing bounded-rational outputs. The prompt is part of the inference context, and the output remains a retrieval-and-recombination operation over corpus material. The objection therefore demonstrates rather than refutes the point: the LLM is fetching examples of bounded-rational behaviour, not computing under a bound itself. The "prompt for it" objection lands at the level of surface behaviour, not at the level of the computational route the paper's argument is about.

## Reasoning models and the level-k bound

A stronger version of the objection is that recent reasoning-distilled models (o1, R1, DeepSeek-R1) demonstrably compute at test time, allocating variable compute to harder problems and producing intermediate chains a base model cannot. The retrieval framing does not need to deny this. The relevant distinction is not whether the LLM computes but what kind of bound caps its computation.

The cognitive-hierarchy literature [Stahl & Wilson, 1995; Camerer, Ho & Chong, 2004] characterises strategic reasoning as a sequence of best-response operations indexed by level k: level-0 acts naïvely, level-1 best-responds to level-0, and so on, with Nash equilibrium corresponding to $k \to \infty$. Empirically, humans cluster at low k, typically 1 to 2, even in lab settings where ample time and writing materials remove most computational constraints. The behavioural-economics literature attributes the cap to cognitive constraints rather than to raw arithmetic capacity: working-memory limits, affective interference, time pressure, attentional bottlenecks. The human δ has structure matching that catalogue.

LLM reasoning is also bounded, but the bound is different in kind. A reasoning model's accessible k is bounded by compute budget, context length, and training-distribution structure. The model's δ, if any, is the gap between achievable k under those bounds and y_strategic. The four operational signatures in Section 6 discriminate accordingly. A δ that shows opponent-individuation effects, distributional asymmetry consistent with loss aversion, and path-dependence under repetition is the human-shaped δ. A δ that scales with compute budget, shows discontinuities at context-length boundaries, and does not respond to opponent framing is the LLM-shaped δ. The two are empirically distinguishable. Reasoning models therefore do not falsify the prediction; they refine the test.

## Section 5 — When δ matters: peer-signal individuation as a moderator

The previous sections argued that human strategic behaviour decomposes into a classical baseline plus an additive residual δ which captures the consequences of bounded computation, and that LLM inference reaches the classical baseline by retrieval rather than by bounded computation. This section asks a more specific question. Even granting that δ is non-zero on average for human play, is δ uniformly non-zero across all strategic settings, or does its magnitude vary with features of the decision environment? The behavioural-economic literature surveyed in Section 4 consistently links candidate contents of δ to social information processing: peer imitation, social comparison, fairness concerns, and affective response to identifiable counterparties. If this is right, then |δ| should scale with how individuated, rather than aggregated, the peer signals in the decision environment are.

### The prediction

In compact form: |δ| is larger in decision environments where the agent observes specific individual opponents than in environments where the agent observes only aggregate opponent signals. An ultimatum game with a named partner, a repeated prisoner's dilemma with a fixed counterparty, or a trust game with a visible recipient should produce larger δ than a p-beauty contest with an anonymous large group, a price-taking auction with only aggregate market feedback, or a public-goods game with only the group total reported back to participants. The same agent, holding everything else constant, should display smaller δ in the second set than in the first.

The prediction carries a quantitative bound. On a standardised effect-size measure of human deviation from y_strategic, individuated-opponent settings should exceed aggregate-opponent settings by at least Cohen's $d \geq 0.5$, a medium effect, holding game structure and stakes constant. Smaller effects would be consistent with the prediction in direction but would not constitute strong support; larger effects, or a reversal of the direction, would respectively confirm or falsify it.

Two things about this prediction deserve flagging. First, the human-side claim restates a regularity well-established in the behavioural-economics literature on social preferences, namely that other-regarding behaviour scales with opponent identifiability. The contribution of stating it here is not to introduce the regularity but to provide the LLM-comparison framing: the prediction is that the human regularity does not transfer to LLM behaviour, because the LLM is not subject to the bounded computation that produces it. Second, the prediction is empirically testable and is not committed to any specific behavioural mechanism. It would survive the revision of any one item in the catalogue of candidate δ-contents, provided the surviving content remains tied to social information processing in some form. It would be falsified by a finding that |δ| is uncorrelated with peer-signal individuation, or that the relationship runs in the opposite direction.

### Independence from the structural claim

The moderator prediction stated here is logically independent of the structural claim in Section 4. The prediction concerns the size of human-side δ as a function of decision environment, and is testable without any commitment to whether LLMs are or are not capable of producing δ at all. If Section 4 is wrong and LLMs can in fact produce δ through some mechanism not yet identified, the gradient prediction still applies to human play and is still informative about which decision environments most depend on δ. If Section 5 is wrong and δ does not scale with individuation, the structural argument in Section 4 still stands as a claim about LLM inference. The two claims are bundled in this paper because they are theoretically related but should be falsified separately.

### What the existing evidence does and does not say

The evidence collected in Section 2 is broadly consistent with the prediction but is too thin and too heterogeneous to constitute a real test. Direction-of-divergence on canonical outcome variables is reported across most game families, but magnitudes are rarely extractable from published abstracts, and human baselines are drawn from heterogeneous source experiments, so cross-game effect-size comparison is not currently feasible from the published record. A proper test would require within-subject design across at least three games spanning the individuation gradient, with sample sizes calibrated to detect a quarter-standard-deviation difference in δ across games, and with both LLM and human arms run under matched information and incentive conditions. The contest-game gap noted in Section 2 is the highest-leverage cell for such a design. Contests can be configured to vary opponent individuation, by switching between named-opponent and anonymous-aggregate-group conditions, while holding the underlying strategic structure of the contest constant. Section 6 takes up the question of how the prequel's claims should be falsified more broadly.

## Section 6 — Conclusion and falsifying experiments

This paper has argued that the additive form behavioural economics uses to model human strategic behaviour is the mathematical signature of bounded computation, that LLM inference reaches the classical baseline by retrieving worked-out solutions from the training corpus rather than by passing through that computational bound, and that the magnitude of the resulting human-side $\delta$ should scale with how individuated the peer signals in the decision environment are. The first two claims are bundled, in the sense that the structural argument in Section 4 depends on the additive form posited in Section 3. The third claim, developed in Section 5, is logically independent of the first two and should be falsified separately. This section operationalises what "structure of $\delta$" means, sketches the design that would falsify the structural claim, and identifies the experimental paradigms most exposed to the moderator prediction.

## Operationalising the structure of δ

For $\delta$ to be a substantive theoretical object rather than a relabelled residual, it must have measurable structure. Four operational signatures discriminate structured $\delta$ from independent noise. Conditional dependence: regress $\delta$ on observable game-environment features (game family, opponent identifiability, stake size, framing), and test the null that $\delta$ is independent of all such features; rejection at $p < 0.05$ with adequate power against medium effect sizes indicates $\delta$ varies with context in a non-noise way. Distributional asymmetry: skewness of $\delta$ in the direction predicted by a candidate behavioural mechanism, with $|\text{skewness}(\delta)| > 0.5$ in the predicted direction; for loss-aversion-tied content, $\delta$ should be left-skewed around the classical baseline in gain frames and right-skewed in loss frames. Path-dependence in repeated games: in human play, $y_t$ should show a non-zero coefficient on lagged $y_{t-1}$ after controlling for game state, while a Granger-causality test on LLM play in the same paradigm should not reject the null. Paraphrase-robustness: present the same game in five or more semantically-equivalent prompt paraphrases; human $\delta$ should remain stable across paraphrases (coefficient of variation below 0.2 on the standardised metric), while corpus-retrieval-mimicked output should show paraphrase-sensitivity. A finding that none of the four signatures holds in a given dataset would constitute evidence that $\delta$ in that dataset is unstructured noise, falsifying the substantive content of the hypothesis.

## Falsifying the structural claim

The structural claim has two falsification routes. The first is on canonical games whose classical-rational solutions are present in standard training corpora. There an LLM that produces structured $\delta$ matching the operational signatures above, without prompted invitation, would refute the claim that retrieval bypasses the bound. The design must measure $\delta_{LLM}$ under a baseline prompt that does not invite behavioural output, then check the residual against all four signatures. Surface mimicry of behavioural output under invitation-prompts does not count, because Section 4.4 already accounts for it as retrieval-and-recombination.

The second route is novel games whose classical-rational solutions are not present in any standard training corpus. There the LLM must compute rather than retrieve, and the prediction is that $\delta_{LLM}$ in such cases is non-zero with structure of its own, derived from training-distribution patterns. The novel-game $\delta_{LLM}$ should therefore look distinguishable from human $\delta$ on the four signatures: in particular, it should not show the path-dependence pattern that bounded human cognition produces in repeated play, and its conditional-dependence structure should track training-distribution game-similarity rather than peer-signal individuation. A finding that novel-game $\delta_LLM$ looks structurally similar to human $\delta$ would refute the retrieval mechanism as the source of the LLM-human asymmetry. To prevent post-hoc re-attribution of any positive LLM $\delta$ to undisclosed training-corpus contamination, novel-game certification must be pre-committed: the game's specific payoff structure and naming conventions should be verified absent from major public training datasets (Common Crawl, arXiv, Stack Exchange, GitHub repositories of game-theory courseware) via targeted search prior to LLM testing.

### Falsifying the moderator prediction

The moderator prediction is falsifiable on a separate track. A within-subject design across at least three games spanning the individuation gradient, with sample sizes calibrated to detect Cohen's $d \geq 0.5$ differences in $\delta$ across games and with matched information and incentive conditions between LLM and human arms, would deliver a real test. The contest-game gap noted in Section 2 is the highest-leverage cell for this design, because contests can be configured to vary opponent individuation while holding the strategic structure of the game constant. All-pay auctions, Tullock contests, and repeated peer-comparison games are the paradigms most exposed to the prediction, in the sense that small changes in opponent-information structure should produce visible changes in human $\delta$-magnitude without changing $y_strategic$. None of these paradigms currently has a published LLM–human contrast.

### Closing note

The prequel is written ahead of empirical follow-up. Stating the framework first allows the follow-up to be interpreted against a defined hypothesis rather than retrofitted to one. The arrangement is the same as it has often been in behavioural economics, where the theoretical statement precedes the experimental programme it then disciplines.